\documentclass[manuscript]{acmart}

\AtBeginDocument{%
  \providecommand\BibTeX{{%
    \normalfont B\kern-0.5em{\scshape i\kern-0.25em b}\kern-0.8em\TeX}}}

\copyrightyear{2023}
\acmYear{2023}
\setcopyright{rightsretained}
\acmConference[FAccT '23]{2023 ACM Conference on Fairness, Accountability, and Transparency}{June 12--15, 2023}{Chicago, IL, USA}
\acmBooktitle{2023 ACM Conference on Fairness, Accountability, and Transparency (FAccT '23), June 12--15, 2023, Chicago, IL, USA}
\acmDOI{10.1145/3593013.3594054}
\acmISBN{979-8-4007-0192-4/23/06}

\usepackage{caption}
\usepackage{subcaption}
\usepackage{pdfpages}

\usepackage{tikz}
\usepackage{xspace}

\usepackage{xcolor}
\usepackage{ifthen}
\newboolean{include-notes}
\setboolean{include-notes}{true}
\newcommand{\nb}[3]{\ifthenelse{\boolean{include-notes}}{{\colorbox{#2}{\bfseries\sffamily\scriptsize\textcolor{white}{#1}}}{\ \textcolor{#2}{\sf\small\textit{#3}}}}{}}

\newcommand{\amsalgorithm}{\textit{AMS algorithm}\xspace}

\makeatletter
\def\NAT@spacechar{~}
\makeatother

\begin{document}


\title{On the Impact of Explanations on Understanding of Algorithmic Decision-Making}

\author{Timothée Schmude}
\email{timothee.schmude@univie.ac.at}
\affiliation{%
  \institution{University of Vienna, Faculty of Computer Science, Research Network Data Science, UniVie Doctoral School Computer Science DoCS}
  \streetaddress{Währinger Straße 29}
  \city{Vienna}
  \state{Vienna}
  \country{Austria}
  \postcode{1090}
}

\author{Laura Koesten}
\email{laura.koesten@univie.ac.at}
\affiliation{%
  \institution{University of Vienna, Faculty of Computer Science, Research Group Visualization and Data Analysis}
  \streetaddress{Sensengasse 6}
  \city{Vienna}
  \state{Vienna}
  \country{Austria}
  \postcode{1090}
}

\author{Torsten Möller}
\email{torsten.moeller@univie.ac.at}
\affiliation{%
  \institution{University of Vienna, Faculty of Computer Science, Research Network Data Science, Research Group Visualization and Data Analysis}
  \streetaddress{Sensengasse 6}
  \city{Vienna}
  \state{Vienna}
  \country{Austria}
  \postcode{1090}
}

\author{Sebastian Tschiatschek}
\email{sebastian.tschiatschek@univie.ac.at}
\affiliation{%
  \institution{University of Vienna, Faculty of Computer Science, Research Network Data Science, Research Group Data Mining and Machine Learning}
  \streetaddress{Währinger Straße 29}
  \city{Vienna}
  \state{Vienna}
  \country{Austria}
  \postcode{1090}
}

\renewcommand{\shortauthors}{Schmude et al.}

\begin{abstract}
  Ethical principles for algorithms are gaining importance as more and more stakeholders are affected by "high-risk" algorithmic decision-making (ADM) systems. \textit{Understanding} how these systems work enables stakeholders to make informed decisions and to assess the systems' adherence to ethical values. Explanations are a promising way to create understanding, but current explainable artificial intelligence (XAI) research does not always consider existent theories on how understanding is formed and evaluated. In this work, we aim to contribute to a better understanding of understanding by conducting a qualitative task-based study with 30 participants, including users and affected stakeholders. We use three explanation modalities (textual, dialogue, and interactive) to explain a "high-risk" ADM system to participants and analyse their responses both inductively and deductively, using the "six facets of understanding" framework by Wiggins \& McTighe~\cite{wiggins_understanding_2005}. Our findings indicate that the "six facets" framework is a promising approach to analyse participants' thought processes in understanding, providing categories for both rational and emotional understanding. We further introduce the "dialogue" modality as a valid explanation approach to increase participant engagement and interaction with the "explainer", allowing for more insight into their understanding in the process. Our analysis further suggests that individuality in understanding affects participants' perceptions of algorithmic fairness, demonstrating the interdependence between understanding and ADM assessment that previous studies have outlined. We posit that drawing from theories on learning and understanding like the "six facets" and leveraging explanation modalities can guide XAI research to better suit explanations to learning processes of individuals and consequently enable their assessment of ethical values of ADM systems. 
\end{abstract}




\maketitle


\section{Motivation}
\label{sec:introduction}

"Algorithmic decision-making" (ADM) systems analyse data to derive information used to support or facilitate decisions~\cite{european_parliament_understanding_ADM_2019}. As such, they are increasingly used in public institutions and administration and thus affect our daily lives.   
Examples include systems for recidivism prediction in criminal justice~\cite{chouldechova2017}, refugee resettlement advice in immigration policy~\cite{Bansak2018}, and employability estimation in public employment~\cite{scott_algorithmic_2022, Allhutter2020en}. The EU classifies ADM systems that are used to decide over human individuals as "high-risk" and proposes to regulate them strictly~\cite{veale_fairness_2018, european_commission_laying_2021}, for example by prescribing adherence to standards of "trustworthy artificial intelligence" (TAI)~\cite{high_level_expert_group_on_ai_eu}. These standards state that a system should be, among other criteria, \textit{transparent}, \textit{fair}, \textit{accountable}, and have \textit{human oversight}, in order to be deemed "trustworthy". However, two problems pose a challenge in fulfilling these criteria: First, no definition is given of when a system is, for example, transparent or fair, and second, a system \textit{can be} transparent or fair, without being \textit{perceived} as such~\cite{lee_algorithmic_mediation_2017}.

How individuals perceive a system's ethical values depends not only on the system's characteristics, but also on individual factors, such as personality traits and demographics~\cite{shulner-tal_enhancing_2022, pierson_demographics_2018}, as well as on the relation between the individual stakeholder and the ADM system~\cite{langer_what_2021, speith_review_2022,jakesch_how_2022}. 
Stakeholders that are involved in a systems' development, deployment, day-to-day usage, or regulation are known to have very different information needs and priorities in assessing ADM systems~\cite{jakesch_how_2022, lee_webuildai_2019, arrietta2020, langer_what_2021}. For example, while an ADM system can produce benefits for an employer, such as informing employee decisions~\cite{Bansak2018} and reducing costs~\cite{marabelli2019}, the same system can negatively impact stakeholders that are the decision targets by discriminating against certain population groups~\cite{woodruff_qualitative_2018, brown_toward_2019}, thus creating two divergent perspectives.

Explanations can aid different stakeholders in acquiring a basic \textit{understanding} of ADM systems in order to "assess" them in terms of ethical values~\cite{langer_what_2021, speith_review_2022}. To this end, numerous studies in Explainable Artificial Intelligence (XAI) examine how people's understanding of a system can be increased by using e.g., input influence, sensitivity, counterfactuals, case-based, and white box explanations~\cite{dodge_explaining_2019, szymanski_visual_2021, shen_designing_2020, cheng_explaining_2019, wang21}. Further, understanding can be affected by the \textit{explanation modality}, meaning the presentation of information in e.g., textual, visual, and interactive form~\cite{szymanski_visual_2021, cheng_explaining_2019}. How to create an explanation that will address every stakeholder's individual information needs, however, remains an open challenge.~\cite{shulner-tal_enhancing_2022, dodge_explaining_2019}.

To acquire a concept of the mental processes involved in understanding, we employ definitions that are established and used in the learning sciences research. Wiggins \& McTighe~\cite{wiggins_understanding_2005}, by referring to Bloom's widely-known "taxonomy of educational objectives"~\cite{bloom_taxonomy_1956, anderson_taxonomy_2001}, suggest that understanding is essentially \textit{transfer}: "to take what we know and use it creatively, flexibly, fluently, in different settings or problems, on our own". Students can demonstrate their understanding by showing their ability to \textit{perform} specific things with their knowledge, which Wiggins \& McTighe~\cite{wiggins_understanding_2005} describe as the "six facets of understanding". Novel explanation methods could benefit significantly from adopting theories such as the "six facets" framework from the learning sciences in order to better construct and evaluate explanations along existent conceptualisations of understanding and learning. Recent XAI studies already began to leverage different theories of understanding to this end~\cite{kawakami_towards_2022, kaur_sensible_2022}.

In this paper, we follow up on these approaches by investigating how a one-on-one explanation presented in three modalities (textual, dialogue, an interactive) creates understanding in different individuals. To this end, we conducted a task-based qualitative study with 30 participants. We analyse their responses using both inductive and deductive approaches, leveraging the "six facets of understanding"~\cite{wiggins_understanding_2005} to examine if participants can \textit{explain}, \textit{interpret}, \textit{apply}, \textit{empathise}, \textit{take perspective}, and \textit{self-reflect} after receiving the explanation. We further provide a practical analysis of the assumption that understanding is a prerequisite for ethical assessment~\cite{langer_what_2021, speith_review_2022}, by relating participants' ability to discuss the fairness of algorithmic decisions to their understanding. As a case study for high-risk ADM, we use the \amsalgorithm -- a system that was planned to be deployed in Austria to predict job-seekers' employability but which was stopped before its actual implementation~\cite{allhutter_bericht_ams-algorithmus_2020}. The \amsalgorithm represents a high-risk system that incited public discourse when it was planned~\cite{Allhutter2020en} and which generalises to other ADM systems used in "Public Employment Services"~\cite{scott_algorithmic_2022} due to the prevalence of individual scoring based on personal attributes. 

We are guided by the following research question, including three sub-questions:\\

\begin{itemize}
    \item \textbf{How does a "global" explanation using textual, dialogue, and interactive modality impact participants' understanding?}
    \setlength{\itemindent}{1em}
    \item RQ1: Which "facets of understanding" emerge in the responses of participants after receiving the explanation? 
    \item RQ2: How is the explanation modality correlated to understanding?
    \item RQ3: To what degree do participants demonstrate the ability to engage in meaningful discourse about the algorithm, for instance in evaluating the algorithms' fairness with regard to decisions about job-seekers?
\end{itemize}

Our findings demonstrate that the explanation method chosen for this study successfully gives participants the opportunity to articulate their thought processes that underlie their understanding of the algorithm. To capture and evaluate these thought processes, we highlight the utility of leveraging learning sciences frameworks, such as the "facets of understanding"~\cite{wiggins_understanding_2005}, to gain insight on participants' understanding. Pertaining to the study design, we validate the one-on-one explanation and interview setup which allows for the gathering of "evidence based on response processes"~\cite{american_educational_research_association_2014_2014} and for an "interactive dialogue" which can "more fully capture understanding"~\cite{sato_testing_2019}. Lastly, we observe that participants can successfully articulate fairness assessments of the given ADM system after proceeding through the explanation, which however vary in detail and argumentative reasoning depending on participants' understanding. We thus practically illustrate the link between understanding and assessing a system's ethical values as posited conceptually in recent XAI studies~\cite{langer_what_2021, speith_review_2022}.

\section{Related Work}
\label{sec:related_work}
\subsection{Relevance of algorithmic decision-making}

In this work, we focus on the \amsalgorithm, a system that is used for "algorithmic decision-making" (ADM) i.e., processing data to support or drive decisions in a public institution~\cite{veale_fairness_2018, brown_toward_2019, cheng_explaining_2019, wang21}. high-risk ADM systems~\cite{european_commission_laying_2021} are increasingly used throughout all countries and sectors, including the COMPAS\footnote{Correctional Offender Management Profiling for Alternative Sanctions.} model to score defendants' recidivism probability in US courts~\cite{chouldechova2017}, the GeoMatch refugee resettlement algorithm~\cite{Bansak2018}, the Dutch, German, and Austrian classification systems for public employment~\cite{scott_algorithmic_2022}, and systems to decide on child welfare services~\cite{brown_toward_2019}. Many of these systems suffer shortcomings, including unreliability of predictions~\cite{brown_toward_2019}, a lack of transparency~\cite{definelicht2020}, missing stakeholder involvement~\cite{scott_algorithmic_2022}, and biased training data~\cite{chouldechova2017}, resulting in negative effects on larger parts of the population due to ADM. Current literature further shows that ADM systems rarely comply with standards such as "trustworthy artificial intelligence"~\cite{high_level_expert_group_on_ai_eu} or "value-based engineering"~\cite{spiekermann_value-lists_2021} for multiple reasons~\cite{madaio_co-designing_2020, definelicht2020, bell_its_2022, dodge_explaining_2019, balagopalan_road_2022}. This is critical since research into the relationship between the use of high-risk ADM systems and societal values suggests that perceptions of such systems as unequal, untrustworthy, or unjust can erode trust in democratic institutions if a large number of people is affected~\cite{radavoi_impact_2020, Hidalgo2021, floridi_ai4peopleethical_2018, oneil_weapons_2016, hermstruwer_fair_2022, binns_its_2018}. Explanations of ADM systems are seen as one of the possible solutions to these problems, as they in theory lead to more transparency and thus more trustworthy systems~\cite{high_level_expert_group_on_ai_eu}. 

\subsection{Stakeholders and explanation design}

We orient our explanation approach towards the high-level goals of explainable artificial intelligence (XAI), which as a research field is dedicated to "amend the lack of understanding of AI-based systems" to enable different groups of people to assess whether a system's output is, e.g., accurate, fair, just, or beneficent~\cite{speith_review_2022, langer_what_2021}. However, the degree of understanding that explanations produce has been shown to vary depending on \textit{who} the explanation's recipient is. XAI literature defines individuals involved in the development, deployment, regulation, or use of an ADM system as "stakeholders". Stakeholders have different information needs and attitudes depending on their relation to the system~\cite{arrietta2020, langer_what_2021}. For example, a "deployer" might expect an explanation to tell them whether the system can inform employee decisions~\cite{Bansak2018} and reduce costs~\cite{marabelli2019}, while affected stakeholders might expect to learn if the system discriminates against certain population groups~\cite{woodruff_qualitative_2018, brown_toward_2019}. 

We use three different explanation \textit{modalities} to present information: textual, dialogue, and interactive modality. In this we are guided by several works that find that explanation modalities can vary in their impact on understanding~\cite{szymanski_visual_2021, cheng_explaining_2019}. We are further guided by works featuring in-person empirical studies~\cite{scott_algorithmic_2022, peck2019, woodruff_qualitative_2018, shen_designing_2020, lee_algorithmic_mediation_2017, lee_webuildai_2019}, as they enable a direct interaction with the participant and, in our case, the introduction of the dialogue explanation modality, in which information is conveyed verbally. A flow chart serves as the basis for all three explanation modalities, as it can depict the complete algorithmic decision-making process~\cite{kulesza_too_2013}, including the "human-in-the-loop", an individual overseeing the algorithm and an essential factor in many ADM analyses~\cite{bell_its_2022, woodruff_qualitative_2018, dodge_explaining_2019, lee_webuildai_2019}.  Using Speith's~\cite{speith_review_2022} taxonomy of explanation methods, our explanation flowchart can be described as a both result- and functioning-focused, model-specific explanation with a visual output format that aims to globally explain the whole decision-making process~\cite{speith_review_2022}. The three modalities build upon this base form and add information via textual, verbal, and interactive presentation.

\subsection{Building and assessing understanding}

The purpose of an explanation arguably entails producing understanding in the explanation's recipient. Many studies discuss and analyse how explanations can influence participants' understanding \cite{balagopalan_road_2022, shen_designing_2020, szymanski_visual_2021, reader_models_2022, shang_why_2022}, but what \textit{constitutes} understanding and how it can be evaluated is not always discussed. Seeing this as preliminary to our discussion, we will provide a brief description of how understanding is discussed in the learning sciences, before introducing Wiggins' \& McTighe's~\cite{wiggins_understanding_2005} framework on understanding and outlining the "six facets of understanding".

In Anderson's and Krathwohl's~\cite{anderson_taxonomy_2001} revised version of Bloom's ~\cite{bloom_taxonomy_1956} taxonomy of educational objectives, understanding is lined up as one of six categories in the "cognitive process dimension", which is counterposed with the "knowledge dimension" to produce the "cognitive" taxonomy of learning objectives. Bloom's original taxonomy was later complemented by the "affective" and "psychomotor" domains; this separation however was criticised "because it isolates aspects of the same objective -- and nearly every cognitive objective has an affective component"~\cite{anderson_taxonomy_2001}. Wiggins and McTighe's~\cite{wiggins_understanding_2005} framework is based on the revised taxonomy of educational objectives, but focuses on the process of understanding. In this work, we therefore use the definition of understanding given by \citet{wiggins_understanding_2005}.

According to \citet{wiggins_understanding_2005}, when someone truly understands a topic, they can: a) "explain", generalise, and make connections; b) "interpret", translate, or make the subject personal through analogies or anecdotes; c) "apply" or "do" the subject in different contexts; d) "take perspectives" on the topic and see the big picture; e) "empathise" with values that others might find odd and perceive sensitively; and f) "self-reflect" on their own beliefs and habits that shape and impede understanding~\cite{wiggins_understanding_2005}. This list of "understanding facets" aims towards "transferability" of knowledge~\cite{wiggins_understanding_2005}. We utilise Wiggins' \& McTighe's~\cite{wiggins_understanding_2005} framework for multiple reasons: First, it includes both the cognitive and affective domain, as well as a notion of "metacognition"~\cite{schraw_promoting_2006} -- meaning to reflect on one's knowledge and understanding. Second, their framework is well applicable to our study design, allowing us to compare our inductive analysis of understanding in participants' responses with a deductive approach. Third, Kawakami et al.~\cite{kawakami_towards_2022} outline a concept of using this framework to produce "learner-centered" XAI, which we follow up on by transferring the theoretical considerations into an empirical study. Lastly, Wiggins \& McTighe~\cite{wiggins_understanding_2005} also provide a categorisation of "barriers to understanding", which we adapt to our use case, consisting of: i) forgetting, ii) being unable to use what we learn, and iii) not knowing that we do not understand. 

We further base our one-on-one interview study design on studies from the learning sciences, which posit that "evidence based on response processes"~\cite{american_educational_research_association_2014_2014} allows us to observe how understanding emerges in the learner's responses and to distinguish it from knowledge or recall~\cite{bransford_contextual_1972}. We further include a task in our study design where participants are asked to explain the \amsalgorithm in their own words, drawing from Duckworth et al.~\cite{duckworth_tell_2001}, who underline that letting learners explain in their own words provides insight into their understanding.

\subsection{Analysing perceptions of algorithmic fairness}

Fairness is seen as one of the central criteria for "trustworthy AI"~\cite{high_level_expert_group_on_ai_eu} or "ethical AI"~\cite{floridi_ai4peopleethical_2018}. Similar to other high-level criteria, the meaning of fairness as a moral value shifts depending on who is asked~\cite{jakesch_how_2022, woodruff_qualitative_2018, pierson_demographics_2018, shulner-tal_enhancing_2022}, and whether it applies to a human or to a machine~\cite{StarkeChristopher2021, lee_algorithmic_mediation_2017}. In this work, we focus on what Langer et al.~\cite{langer_what_2021} call the "epistemic" satisfaction of fairness, meaning that we examine whether participants can engage in discourse about the fairness of the \amsalgorithm after receiving our explanation. In contrast, we are \textit{not} focussing on the "substantial" satisfaction of fairness~\cite{langer_what_2021}, meaning our discussion will not cover whether the \amsalgorithm actually acts fairly or not. We thus use the fairness assessment as an indicator of whether the explanation served to increase participants' understanding and enabled them to assess the system in ethical terms. 

\section{Methodology}

To gain insight into stakeholders' understanding of ADM, we conducted a task-based qualitative study with 30 participants using three explanation modalities of the \amsalgorithm (textual, dialogue, and interactive). In the study, an explanation of the algorithm (Section~\ref{sec:treatments}) was followed by two tasks (Section~\ref{sec:tasks}) and a short interview about the deployment of the algorithm in society. We analyse participants' responses inductively and deductively, using the "six facets of understanding" framework~\cite{wiggins_understanding_2005}. An overview of the study procedure is depicted in~\autoref{fig:procedure}. See Section \autoref{fig:Participant_table} for the participant sample.

\subsection{The algorithm}
\label{sec:algorithm}

In our study, we used the \amsalgorithm\footnote{The abbreviation AMS stands for the Public Employment Agency.} as a prototypical example of an algorithmic decision-making system in Public Employment Services~\cite{scott_algorithmic_2022}. The algorithm was developed between 2015 and 2021 by a private company for the Public Employment Agency and was piloted for a short time in the autumn of 2018, but was never used as a live system and is currently put on hold due to legal objections~\cite{allhutter_bericht_ams-algorithmus_2020}. The case has been covered by several academic studies and incited public discourse over the benefits and risks of its deployment~\cite{allhutter_bericht_ams-algorithmus_2020, Allhutter2021de, scott_algorithmic_2022, lopez_reinforcing_2019}.\footnote{An extended discussion of the public discourse and perception of the \amsalgorithm is provided in Section A in the supplementary material.} As a great number of people could be affected by the implementation of such a system, and as the stakes generalise well to other high-risk settings, we use the \amsalgorithm as a case example for our study.

\paragraph{The algorithm's predictions and model}
The \amsalgorithm was constructed to assign job-seekers to one out of four categories ("high", "medium", or "low" employment chances, plus special cases), depending on their personal attributes, such as age, gender, and education. Every prediction would be confirmed or corrected by an employee of the Public Employment Agency. The groups of employability were defined as follows:
\begin{itemize}
    \item "Medium": Job-seekers would receive regular support measures\footnote{Such as further training or application coaching.} to improve their chances of finding employment.
    \item "High": Job-seekers were expected to find new employment quickly and would receive fewer support measures.
    \item "Low": Job-seekers were expected to require more support and would be referred to another facility.
    \item Other: Teenagers, people with disabilities, and people over 50 would receive additional support measures independent of their employability scoring~\cite{allhutter_bericht_ams-algorithmus_2020}.
\end{itemize}

The algorithm's model was trained on several years of job-seekers' data, mainly consisting of personal attributes (features): gender, age, citizenship, education, impairment, obligations of care (only women), occupational group, prior occupations; as well as a representation of the local job market.\footnote{Attributes are listed in detail in Section A in the supplementary material.} People with similar personal attributes would be grouped and compared to the "standard group"~\cite{Holl2018} -- young men with secondary school education -- to be assigned a short-term\footnote{At least 90 days unsupported employment in seven months. Unsupported employment is not subsidised by the Employment Agency.} and long-term\footnote{At least 180 days unsupported employment in 24 months.} employability score. For further details please refer to the supplementary material and Allhutter et al.~\cite{allhutter_bericht_ams-algorithmus_2020}.

\paragraph{Biases in the algorithm}
The weighting of personal features in the algorithm's classification can be considered biased in that several attributes such as gender and nationality led to decreased employability predictions~\cite{Holl2018}. However, these biases would in theory lead to higher support measures for job-seekers possessing these attributes, in effect supporting those that were potentially disadvantaged in the job market~\cite{Holl2018_Standards}. Nonetheless, Allhutter et al.~\cite{Allhutter2020en} point out how the algorithm's practical implementation could have detrimental effects on the support and treament that job-seekers would receive. In summary, the \amsalgorithm is an example of how personal and systemic considerations can lead to value conflicts and ethical dilemmas in algorithmic decision-making, which is why we chose it as a suitable example for the study. Explanation and task examples were chosen such that these issues were brought to the participants' attention, without however preempting any judgement or value statement.

\subsection{Study setup}
\label{sec:setup}

\subsubsection{Explanation modalities}
\label{sec:treatments}

\begin{figure*}[]
    \centering
    \includegraphics[width=\textwidth]{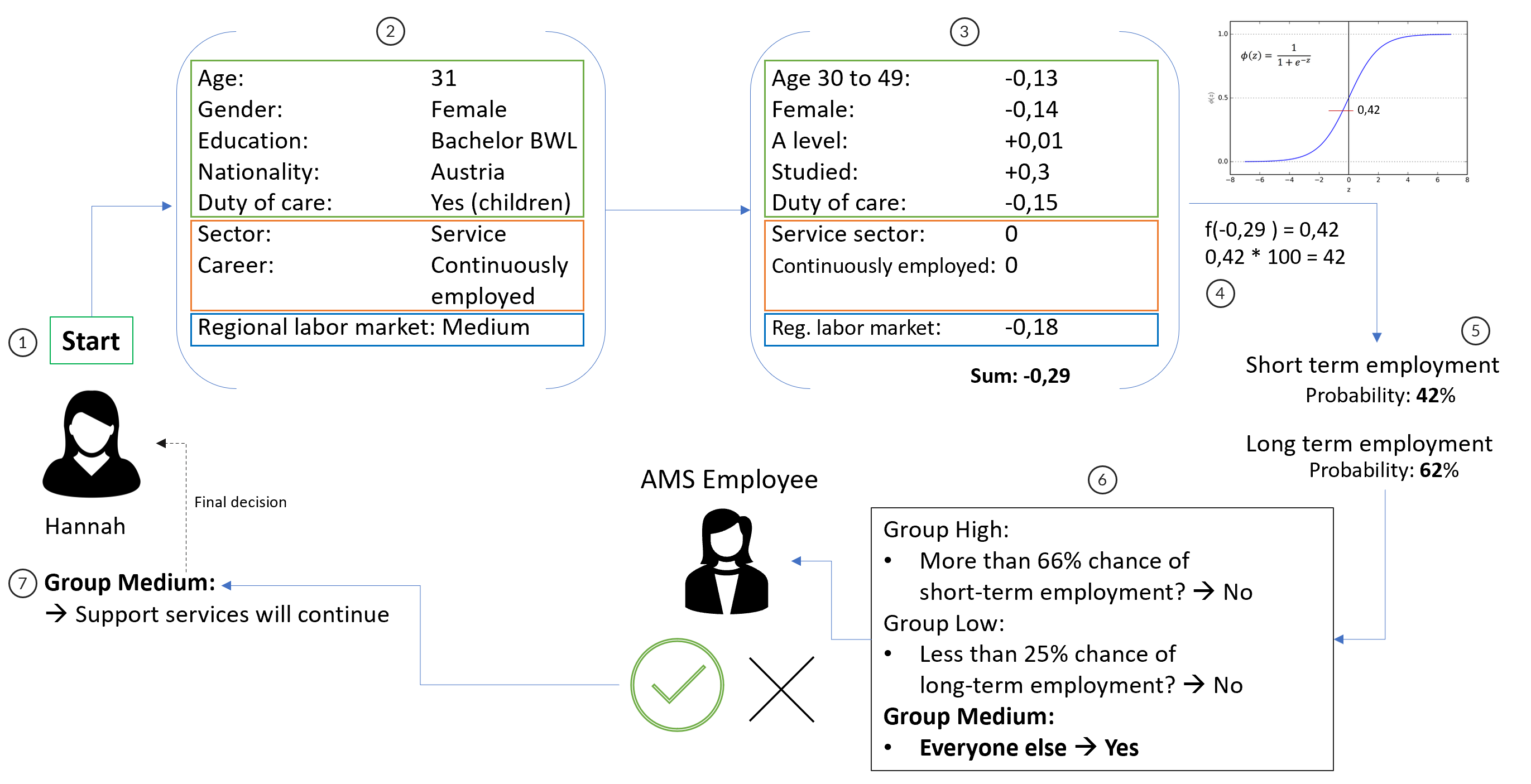}
    \caption[Visual-dialogue explanation]{The flowchart that every explanation modality built upon. The different parts of the explanation are presented sequentially according to the indicated numbering. \textcircled{1} \textit{Hannah} is a fictional job-seeker reporting to the Employment Agency. \textcircled{2} Some of her personal attributes are recorded for the employability prediction. \textcircled{3} \textit{Hannah's} attributes are compared to the "standard group": young men with secondary school education. Based on this comparison, the weight and value of \textit{Hannah}'s attributes are calculated. \textcircled{4} The sum of these values is put into a logistic regression function that maps it to a scale of 0 to 1. Multiplied by 100 this gives the short-term employability chance of \textit{Hannah}. \textcircled{5} The long-term chance is calculated in a similar fashion with a different model. \textcircled{6} According to three simple rules, \textit{Hannah} is assigned to one of three groups, which is then confirmed or corrected by the employee of the agency. \textcircled{7} \textit{Hannah} receives the final decision and group assignment.}
    \label{fig:explanation}
\end{figure*}

\begin{figure*}[]
    \centering
    \includegraphics[width=\textwidth]{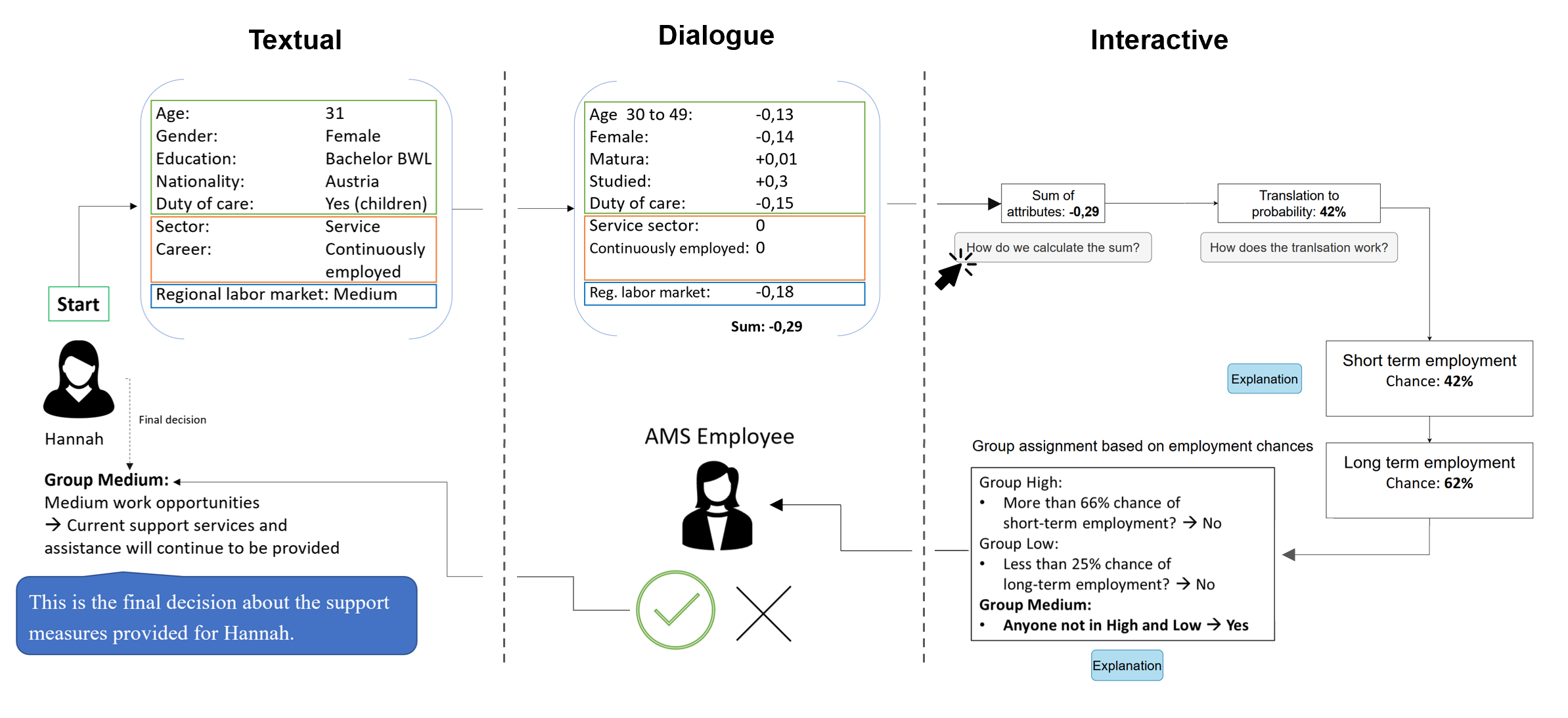}
    \caption[Explanation modalities]{The flowchart split into all three modalities: textual, dialogue, and interactive. Information was added by textual comments, verbal comments, and interactive controls, respectively.}
    \label{fig:explanation_modalities}
\end{figure*}

We used a between-subjects study design, showing each participant one of the three distinct explanation modalities (textual, dialogue, interactive) of the \amsalgorithm. All modalities built upon a visual flowchart of the algorithmic decision-making pipeline, depicted in \autoref{fig:explanation}\footnote{Please note that alt-text for Figures 1 and 2 is provided in Section G of the supplementary material.}. In the explanation, the fictional job-seeker \textit{Hannah} reports to the Employment Agency and is assigned to group "medium" by the algorithm, which is confirmed by the employee. The modalities, depicted in \autoref{fig:explanation_modalities}, enriched the flowchart with additional information, which was identical between modalities but was presented in different ways\footnote{For a more detailed description of modalities confer Section C in the supplementary material.}:

\begin{itemize}
    \item \textbf{Textual}: Textual descriptions were added to the basic flowchart in the form of comments. Participants continued through the explanation slides at their own pace and asked questions at the end.
    \item \textbf{Dialogue}: The study examiner verbally added information to the flowchart, explaining depicted components in each slide. Participants could ask questions throughout the whole process.
    \item \textbf{Interactive}: The flowchart was implemented as a simple interactive web version that featured buttons which showed textual descriptions when clicked. Participants could ask questions at the end.  
\end{itemize}

We chose these modalities to achieve multiple aims: First, showing the whole "global" decision-making process without requiring participants to have technical or specific domain knowledge, following the setups of Wang \& Yin~\cite{wang21} and Logg et al.~\cite{LOGG201990}. Second, giving learners the opportunity to close gaps in their understanding by asking questions ("inquiring") and interacting verbally in the dialogue modality, which can lead to more effective understanding~\cite{sato_testing_2019, robertson_responsive_2015, smith_why_2009, liao2020}. Third, examining the contrast between static and interactive interfaces, motivated by Cheng et al.~\cite{cheng_explaining_2019}, who found that interactive interfaces led to increased understanding.

\subsubsection{Tasks: predicting employability and explaining the algorithm}
\label{sec:tasks}

Participants had to complete two task sections that were meant to probe their understanding and fairness assessment of the algorithm, as depicted in \autoref{fig:procedure}.\footnote{Example cases are described in detail in Section D and E in the supplementary material. Cases were taken from a detailed report on the \amsalgorithm by Allhutter et al.~\cite{allhutter_bericht_ams-algorithmus_2020}.} 

In the first task section, participants were presented with three example cases of job-seekers and were asked to (1) propose measures that could help the person find work again and (2) estimate their chances for short-term and long-term employment\footnote{Short-term in the \amsalgorithm is defined as being employed at least 90 days in the next seven months, long-term as at least six months in the next two years}. Participants then received the algorithmic decision for the job-seeker and the employee's decision (accepting or correcting the algorithm's scoring plus any additional measures), and indicated whether they perceived the (3) algorithmic and (4) human decision as fair. 

The second task section was split into two: In the first sub-task (2.1), participants were provided with a job-seeker case and were asked to explain to the study examiner how the algorithm would handle the case. In the second sub-task (2.2), participants received a case similar to the first one but ranked higher in terms of employability. Participants should then indicate why the two job-seekers were classified differently.\footnote{The first case was female and had duties of care, while the second was male and did not have duties of care (for further detail see supplementary material).} 

\begin{figure*}[]
    \centering
    \includegraphics[width=\textwidth]{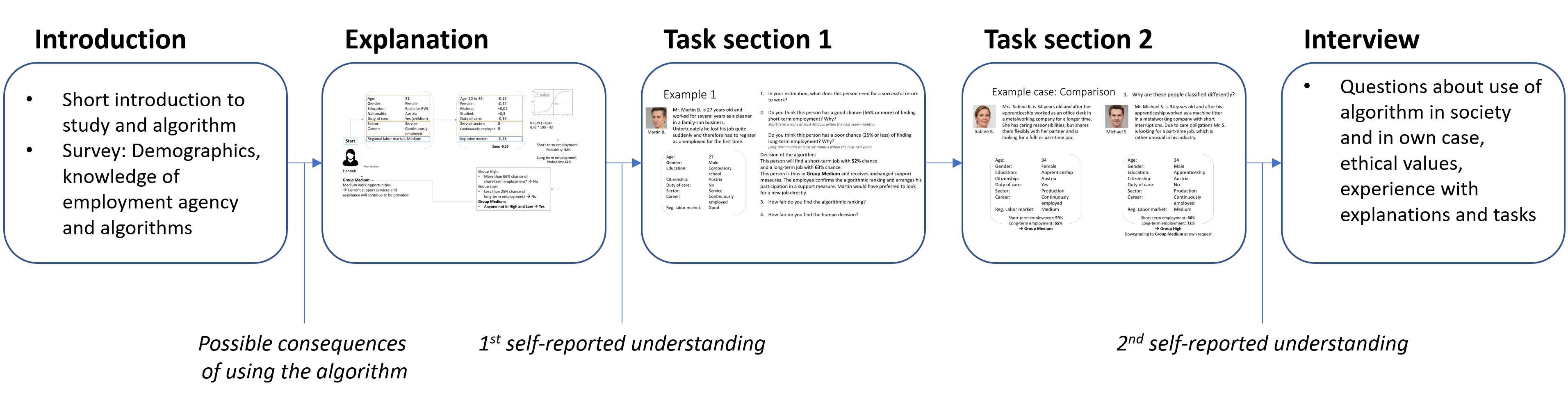}
    \caption[Study procedure]{Depiction of the study procedure. Participants received a short introduction about the goals and scope of the study, filled out a questionnaire, and then were asked about the possible consequences of implementing the \amsalgorithm. They then received one of three explanation modalities and were asked about their self-reported understanding. Participants then proceeded through both task sections, were asked again for their self-reported understanding and finally answered several interview questions.}
    \label{fig:procedure}
\end{figure*}

\subsection{Analysis}

For our thematic analysis, we applied two levels of qualitative coding to the data: inductive analysis in the first pass, and deductive analysis in the second.\footnote{Both approaches are documented in the supplementary material.} 

In our first pass, we examined the data for understanding and perceptions of fairness, letting the overarching themes and theory emerge from the data, following Thomas~\cite{Thomas2006}. We created 18 categories capturing different aspects of understanding, such as the level of detail (e.g., whether participants attended more to the technical details or the societal consequences), the connection to personal knowledge and experience, emotional engagement, and instances in which understanding was impeded. We created 18 more categories to capture participants' perceptions of algorithmic fairness, including statements about the algorithm's precision, the importance of the human in ADM, and issues of inequality. We then compared statements addressing algorithmic fairness with the four dimensions of algorithmic fairness perceptions by \citet{StarkeChristopher2021} to find intersections and create overarching themes.

In our second pass, we produced six deductive code categories according to the "six facets of understanding" framework by \citet{wiggins_understanding_2005}, which we split, according to the sub-processes involved in every category, into a total of 21 codes. We analysed the interviews once more using this deductive framework with the aim to examine whether participants' responses could be mapped to these pre-defined categories and to examine whether the framework captured aspects that our inductive codes missed. We found that while most inductive codes from our first pass could be assigned to one of the "six facets of understanding", the framework also introduced additional differentiations that were useful for the analysis, which will be further described in Section~\ref{sec:findings}. A detailed description of both inductive and deductive codes is further included in the supplementary material.     

\subsection{Participants}
\label{sec:recruting}

In \autoref{fig:Participant_table}, we present a description of the 30 study participants, whom we recruited in four locations: a café near a university, a café in another city district, an auto repair garage, and a local employment agency's office. We conducted three of the studies online; all others were conducted personally in place. In our recruitment, we aimed for the inclusion of the two stakeholder groups "users" and "affected stakeholders", a balance in terms of age, gender, and occupation, as well as a focus on people with no expert knowledge of algorithms. While of course not representative of the general population, the participant sample allowed us to gather a wide range of perspectives on how the \amsalgorithm was understood and perceived in terms of fairness. 

As motivated in Section~\ref{sec:introduction}, algorithmic decision-making systems can be detrimental to democracy if their implementation is faulty or unaware of the public's stance towards the system~\cite{radavoi_impact_2020, Hidalgo2021, floridi_ai4peopleethical_2018, oneil_weapons_2016, binns_its_2018}. For this reason, we aimed to conduct the study with a balanced sample of the "general public", with the only condition being that participants would optimally not be experts on algorithms. The sample included both employees in the public employment sector and job-seekers, but also individuals of different educational and cultural backgrounds who were not directly affected by public employment. While our sample is not representative, it includes a range of different perspectives on the explanation, which provided a solid reasoning ground for examining the explanation's effects on understanding.

\begin{table}[]
\footnotesize
\caption{Details on the participants of our study. 5 participants were employees of the Public Employment Agency or similar institutions, whom we define as "domain experts" and who are indicated with a \textit{d} attached to their ID. 5 participants were job-seeking at the time of the study, whom we define as "affected stakeholders".}
\begin{tabular}{llllll}
\hline
ID  & Explanation & Age & Gender & Education               & Occupation           \\ \hline
1   & textual     & 32  & F      & Master                  & Researcher           \\
2   & textual     & 26  & M      & Master                  & PhD Student          \\
3   & textual     & 23  & D      & Bachelor                & Master's student     \\
4   & textual     & 36  & F      & Master                  & UX   Researcher      \\
5   & textual     & 49  & M      & PhD                     & Financial Advisor    \\
6   & textual     & 25  & M      & 1st state examination & PhD Student          \\
7   & textual     & 47  & F      & Academy                 & Leisure pedagogue    \\
8d  & textual     & 41  & M      & Academy                 & Social worker        \\
9   & dialogue     & 26  & M      & Bachelor                & Student              \\
10  & textual     & 24  & M      & Apprenticeship          & Car mechanic         \\
11  & textual    & 24  & M      & Apprenticeship          & Car mechanic         \\
12  & dialogue    & 26  & M      & Apprenticeship          & Car mechanic         \\
13  & dialogue    & 58  & M      & Bachelor                & Clerk                \\
14  & dialogue    & 69  & F      & University              & Pension              \\
15  & dialogue    & 42  & F      & University              & Job-seeking          \\
16  & dialogue    & 52  & F      & Apprenticeship          & Job-seeking          \\
17  & dialogue    & 51  & M      & Vocational college      & Job-seeking          \\
18d & dialogue    & 60  & F      & Master                  & Application trainer  \\
19d & dialogue    & 57  & M      & A level                 & Personnel consultant \\
20d & dialogue    & 55  & F      & University              & Personnel consultant \\
21  & interactive & 28  & F      & Bachelor                & Student              \\
22  & interactive & 23  & F      & A   level               & Student              \\
23  & interactive & 37  & M      & University              & Employed             \\
24d & interactive & 48  & M      & University              & Trainer              \\
25  & interactive & 26  & F      & Apprenticeship          & Short-term worker    \\
26  & interactive & 60  & M      & University              & Job-seeking          \\
27  & interactive & 35  & M      & University              & Self-employed        \\
28  & interactive & 27  & M      & Master                  & Journalist           \\
29  & interactive & 29  & F      & Master                  & Consultant           \\
30  & interactive & 31  & M      & Bachelor                & Consultant           \\ \hline
\end{tabular}
\label{fig:Participant_table}
\end{table}

\section{Findings}
\label{sec:findings}

In this section, we present the findings from our study, structured according to our RQs covering the following topics: understanding (Section~\ref{sec:rq1}), the effect of explanation modality (Section~\ref{sec:rq2}), and enabling the ethical assessment of ADM (Section~\ref{sec:rq3}).\footnote{Please note that some participant responses touch on sensitive topics, such as discrimination and self-harm.}

\subsection{Which "facets of understanding" emerge in the responses of participants after receiving the explanation? (RQ1)}
\label{sec:rq1}

\subsubsection{"Emotional" and "analytical" facets of understanding:} Participants' understanding of the explanation surfaced in different ways. The framework of Wiggins \& McTighe~\cite{wiggins_understanding_2005} contains both "analytical" (explain, apply, take perspective) and "emotional" (interpret, empathise, self-reflect) facets of understanding. Participants tended to show facets of either one of the two sides, and the degree to which they felt personally affected by the algorithm made a difference in which facets they were most likely to use.
For example, participants who were looking for employment at the time of the study rather responded to the question of whether the \amsalgorithm should be used to score job-seekers with the facet "empathise":

\begin{quote}
    \textit{These are just numbers, you don't know the background of why they lost the job. It's not fair because it's just a program and the human side is completely gone.} (P17)
\end{quote}

In contrast, participants who never had contact with the Public Employment Agency and did not know anyone who was officially unemployed responded to the same question rather by critically "taking perspective" on the algorithm's application:

\begin{quote}
    \textit{I think it's good as a recommendation. I think it turns into a procedure that is otherwise purely a human decision and provides a percentage probability, though I wouldn't use it on its own. I think it's good because the case worker has something to go by, but for some examples, it's not complex enough, that's what the human case worker is for.} (P6)
\end{quote}

\subsubsection{Relating to the information in more than one way indicates higher understanding}

According to Wiggins \& McTighe~\cite{wiggins_understanding_2005}, successful understanding must show \textit{all} of the six facets: "explain," "interpret," "apply," "empathise," "take perspective," and "self-reflect." While no participant used all six facets, we observe that several participants combined up to four facets in a single answer. In the following quote, the participant "explains" the current labour market situation, "interprets" what this would mean for the job-seeker, and as a result "takes perspective" by critically questioning the algorithm's employability scoring:

\begin{quote}
    \textit{I disagree with the algorithm in this case because now there are shortages of employees in specific sectors like catering. So employers are forced to bend their requirements and will have a positive attitude towards applications from people such as Harald. He has a lot of experience, and he is motivated, so they will disregard his age, which would be bad in normal circumstances.} (P23)
\end{quote}

Notably, neither stakeholder group, demographic background, nor prior knowledge seemed to affect if someone would show the ability to use multiple facets of understanding to make sense of the algorithm. This stands in contrast to previous studies, which found that measurable understanding increases with domain expertise~\cite{cheng_explaining_2019, szymanski_visual_2021} and the level of education\cite{shulner-tal_enhancing_2022}. 

\subsubsection{High understanding co-occurs with self-reflection}

Participants that used multiple facets of understanding and thus showed a higher understanding were able to give comprehensive answers to the tasks and often supplemented their responses by reflecting on their own understanding, beliefs, or circumstances. This occurred, for example, in response to the idea of being classified by the \amsalgorithm:

\begin{quote}
    \textit{In my case it would not be a disadvantage, because I am privileged in terms of education and I have no care obligations. [...] But if I were less privileged I wouldn't want that and if I had support needs I wouldn't want that either.} (P1)
\end{quote}

Reflections tended to occur at a "turning point" of the interview, consisting of a question asking if the algorithm should be used on everyone, followed by a question asking if it should be used in the participant's own case. Often when participants responded to these questions disparately, for example by voting against the general usage, but agreeing to the usage on themselves, they felt inclined to reflect on their understanding and their beliefs without being prompted.

\subsubsection{Barriers to understanding} 

Wiggins \& McTighe~\cite{wiggins_understanding_2005} define three cognitive processes that can hamper understanding and learning: i) forgetting, ii) being unable to use what we learn, and iii) not knowing that we do not understand. Forgetting was an issue distributed throughout the participant sample, as participants could not refer back to the explanation while proceeding through the tasks. Notably, participants with lower levels of education encountered more issues of forgetting and were not always able to apply the learned information, which found its expression in incorrect recollections of the explanation and less emergence of understanding facets: 

\begin{quote}
     \textit{I didn't understand the conversion, I don't know the formula. Otherwise, I halfway understood it, they simply take the data and... I don't know how to explain it.} (P10)
\end{quote}

In contrast, we did not find many instances of a participant not realising their lack of understanding. This, however, is more likely due to the difficulty of distinguishing it from the inability to apply knowledge. Finding a way to better identify this barrier to understanding could be a topic for future work.   

\subsubsection{Facets not covered by the framework}

The comparison between inductive and deductive thematic analysis shows that while most inductive themes of understanding could be mapped to the theoretical framework of Wiggins \& McTighe~\cite{wiggins_understanding_2005}, we also identified themes that the framework did not include. First, the "six facets of understanding" do not provide a clear categorisation of expressions that reflect a strong value statement, such as "I believe that algorithms actually have no place in this field." (P18d). The closest match is the facet "interpret," which however aims more towards making the topic "personal or accessible through images, anecdotes,
analogies"~\cite{wiggins_understanding_2005}, and less towards value statements. It is debatable whether these statements bear evidence for understanding, but seeing that Langer et al.~\cite{langer_what_2021} describe value statements as an indicator for stakeholder understanding needs, this lack of categorisation stood out.

Second, our application of the framework does not allow us to distinguish between different domains of knowledge, which in our analysis sometimes blurred the lines between understanding the algorithm and understanding the public employment system. A clearer distinction between the subjects of understanding would be useful for future analyses.

\subsection{Which, if any, correlation to the explanation modality can be seen in the understanding facets? (RQ2)}
\label{sec:rq2}

The different modalities of how the explanation was presented (textual, dialogue, and interactive) had a limited effect on the emergence of specific understanding facets. However, we point out two findings related to the textual and dialogue explanation modality.

\subsubsection{Textual modality leads to understanding barriers} The "textual" modality group showed more and deeper barriers to understanding than other groups. Several participants commented on the amount of text in the explanation and their learning process:

\begin{quote}
    \textit{I have to be honest, I find it easier when someone explains something to me. Then I can also ask until I get it. It is hard for me to understand something like that just by reading it.} (P10)
\end{quote}

In contrast, participants expressed their satisfaction both with the dialogue and interactive modality, praising the option to ask questions and the engagement with the explanation. Of all 20 participants in the dialogue and interactive modality groups, only one stated that they would have rather liked a textual explanation. 

\subsubsection{Dialogue modality leads to increased expression of some understanding facets} Compared to the textual and interactive modality, the dialogue explanation showed an overall increase in the number of words spoken by participants and more usage of the facets "interpret", "empathise", and "self-reflect". Factual questions by participants mostly served the purpose of confirming information that was already present in the flowchart (e.g., "\textit{So, having studied actually has the largest effect on your score?}" (P13)). In contrast, questions that pertained to personal concerns or systemic issues often touched on the "emotional" facets of understanding  (e.g., "\textit{Do women automatically get less points? Where does discrimination begin, where does it end? (P15)}"). These questions were not answered in detail during the explanation, but instead served as points of entry to the later interview. The dialogue modality might have thus helped to later start the in-depth conversation about the algorithm by introducing a verbal interaction directly at the beginning and giving participants more opportunities to correct their understanding. Other possible reasons for these findings will be discussed in Section \ref{sec:discussion}.

\subsection{Do participants demonstrate the ability to engage in meaningful discourse about the algorithm, for instance in evaluating the algorithms' fairness with regard to decisions about job-seekers? (RQ3)}
\label{sec:rq3}

After each case example in task section 1, participants were asked to assess and discuss the fairness of both the algorithmic and human decisions regarding the case. We used these fairness assessments as a form of proxy for their ability to engage in discourse about the algorithm's ethical dimensions. We find that all participants were able to articulate a basic fairness assessment, but that their statements differed strongly in detail and argumentative reasoning. 

\subsubsection{Participants demonstrate the ability to individually assess algorithmic fairness}

The "epistemic" satisfaction of a trustworthy AI criterion means that people are able to assess on their own grounds whether a system fulfills a certain ethical criterion, such as being fair or not~\cite{langer_what_2021}. This does not mean that the system \textit{is} fair, only that people are able to \textit{discuss} it. Participants fulfilled this "epistemic" criterion as they showed a rich and diverse range of fairness assessments of the \amsalgorithm, including topics such as the influence of the "human-in-the-loop", the perceived gender inequality, the perception of "algorithmic objectivity" and, in parallel to findings from Scott et al.~\cite{scott_algorithmic_2022}, the importance of using the system for "orientation purposes, not to deny access to resources". 

We further observe that the fairness assessments changed depending on whether participants made more use of analytical or emotional facets of understanding, as evident in these responses to the algorithm predicting "low" employability for a job-seeker:

\begin{quote}
    \textit{I see him almost like me. He's 49, so for me, he would already count as 50+ and should get extra support. [...] \\I find the algorithm good and the human decision bad, because it goes by numbers and is not accommodating. The algorithm just sees 49 and makes the decision.} (P17)
\end{quote}

\begin{quote}
      \textit{I understand why the algorithm says Group "Low" in this case. If he retrains and can explain his career well, I think he could find work, but needs support to do so. [...] Personally, I would perhaps not rate him that way, but I think it's good that he gets into the group.} (P6)
\end{quote}

The first participant relates the decision to his own personal circumstances and speaks about the difference between human and algorithmic decision-making, while the second takes a more analytical approach in speaking about the consequences of the decision to justify his assessment. Despite their different reasons, both participants were able to argue why they agreed with the algorithm, thus fulfilling the "epistemic" criterion~\cite{langer_what_2021}. At the same time, this case exemplifies that individuality in understanding also leads to individuality in the fairness assessment.  

\subsubsection{Understanding barriers lead to less nuanced fairness assessments} As expected, participants who encountered more understanding barriers assessed the algorithm's fairness in much less detail. Often, some form of blanket statement was used that addressed neither dimensions of fairness nor facets of understanding:

\begin{quote}
    \textit{I think the algorithmic classification is fair, because I was of the same opinion.} (P10)
\end{quote}

The difference to more detailed fairness assessments becomes apparent when we compare this to statements from participants with the same demographics, mostly similar education, but a different explanation modality and largely no barriers to understanding:

\begin{quote}
    \textit{I'll try to leave out the current situation in the labour market. I mean, he has his problems. I understand that the chance is low, but this is too low, the algorithm and especially the employee are not fair. Perhaps the employee is new and has no experience in how to place people in the labour market.} (P12)
\end{quote}

The latter statement shows three different facets of understanding while the participant reasons about the fairness assessment: taking a critical perspective on the algorithm, reflecting on the particularities of the post-COVID labour market, and including the perspective of the employee.

\subsubsection{Unwillingness to engage in discourse about the algorithm's fairness}

Besides understanding barriers, the most common reason for a lack of fairness assessments was a general objection to assessing the system as "fair" or "unfair". These objections were often accompanied by the (flawed) argument that an "objective" algorithm was not able to act fairly or unfairly:

\begin{quote}
    \textit{What does fair mean? I can understand the algorithmic decision. Fair is such a strong moral word, I don't find it immoral, I rather find it "justified", "understandable", "realistic".} (P29)
\end{quote}

Similar responses were given when participants were prompted to rate the algorithm as "just", "legitimate", "social", "biased", and "democratic". Several participants commented on the close meaning between words, which points to the need of differentiating concepts when asking for complex moral value judgements. 

\section{Discussion}
\label{sec:discussion}

In this section, we discuss our findings with regard to the research questions and suggest directions for designing explanations that create increased understanding in different stakeholders and enable the ethical assessment of ADM. 

\subsection{Designing explanations to address all six facets of understanding} The "understanding by design" framework~\cite{wiggins_understanding_2005} states that the primary objectives in creating understanding are to convey a topic's central ideas, address all "facets" of understanding, and uncover misunderstandings. Applying the framework to our explanation setup allowed us to gather insights into the participants' mental processes involved in understanding, i.e., which facets emerged for which participant and how many facets emerged simultaneously. 
In particular, we want to highlight that participants showed "emotional" understanding facets, such as "empathise" and "interpret", which are not always addressed in ADM explanations, despite their known importance in human perception of ADM systems~\cite{woodruff_qualitative_2018, scott_algorithmic_2022, lee_algorithmic_mediation_2017}. The facet "self-reflect" is another valuable dimension that could improve future explanations, as several studies suggest that "metacognition", in the form of reflecting on one's knowledge and understanding, is a "most powerful predictor of learning"~\cite{veenman_metacognition_2006, schraw_promoting_2006}.

Theories on learning and understanding like the "six facets"~\cite{wiggins_understanding_2005} can thus guide explanation design by identifying distinct learning goals and by outlining mental processes that support understanding. We chose Wiggins' \& McTighes'~\cite{wiggins_understanding_2005} framework due to the practical interpretation of understanding and the well-established theoretical foundation in Bloom's taxonomy~\cite{bloom_taxonomy_1956, anderson_taxonomy_2001}. Other promising sources for future studies include the concept of "responsive teaching"~\cite{robertson_responsive_2015}, "knowledge building and knowledge creation"~\cite{scardamalia_knowledge_2014}, and "sensemaking theory"~\cite{weick2005}. We posit that drawing from these theories in the design and development of explanations can guide research to better support "humans in learning about particular AI systems and how to work with or around
them"~\cite{kawakami_towards_2022}.  

\subsection{Effect of explanation modality on understanding} Our findings suggest that the textual and dialogue modality had the most effect on understanding. The textual modality led to less emergence of understanding facets for several participants, some of whom stated that they did not usually rely on text to learn information. Compared to Szymanski et al.~\cite{szymanski_visual_2021}, where participants disliked textual information but actually performed better using them, participants in our case did not show any advantages in understanding after receiving the textual explanation. 

The dialogue modality, in contrast, led to an overall increase in observed facets of understanding, which could have multiple reasons: i) the higher amount of words spoken, ii) the option to ask questions during the explanation, and iii) the additional personal interaction. Although the dialogue modality in theory allowed for a higher amount of factual input, participants seldom chose to ask more than four or five brief factual questions, which mostly pertained to secondary details of the algorithm (e.g., the weighting of certain features) and in terms of information differed little from the textual modality. At the same time, several participants used the dialogue modality to express their opinions and attitudes towards specific information and to ask more profound questions about e.g., the intention behind the algorithm's deployment and the selection of the "standard group". The dialogue modality might thus serve as a conversational ice-breaker due to the direct interaction between "explainer" and participant, encouraging participants to share their own thoughts and, in turn, increase their understanding. This also connects to findings from Miller~\cite{miller_explanation_2019}, who states that explanations between humans are "social" and "presented as part of a conversation." This is striking considering that verbal or dialogue explanations are seldom used in contemporary XAI research. Our findings establish this explanation modality as a valid alternative to be explored in future ADM explanations. 

\subsection{Enabling participants to engage in discourse about the algorithm's ethical values} We used the fairness assessment of the \amsalgorithm as a form of proxy to observe whether people are able to articulate value assessments after receiving the explanation. We find that i) the explanation provided most participants with the necessary information to form a detailed fairness assessment, and thus enabled them to successfully engage in discourse about the algorithm’s ethical values, and ii) in contrast to other participants, domain experts were able to discuss the algorithm’s implementation even before the explanation, stating that they had come in contact with automated tools.
On the other hand, participants who showed barriers to understanding articulated less nuanced fairness assessments. Further, some participants stated that an algorithm simply could not be judged in terms of fairness, despite showing multiple facets of understanding in their responses. This means that understanding the system might not be the only prerequisite for stakeholders to make a value assessment, but that understanding what specific ethical values mean when being applied to the system might be just as important. Future explanations should thus consider adding an explanation of the ethical values or finding stand-ins that convey the core of the value in a more accessible manner.        

\subsection{Limitations} In this section, we touch on four limitations of our study. First, our assessment of understanding relied largely on qualitative analysis of participants' responses, which were self-reported and thus might limit objectivity. For future studies, we therefore consider including a quantitative evaluation of understanding. Second, some inductive codes such as strong value statements were not covered in the "six facets" framework~\cite{wiggins_understanding_2005}. Seeing that value statements can "serve as an orientation for when stakeholders are more likely to demand higher degrees of understanding"~\cite{langer_what_2021}, future analyses should consider including a facet covering these forms of statements. Third, the study design relied on memorisation of information, as no reference was given to participants while proceeding through the tasks. While this impacted some participants, it also highlighted a difference in memorability of the textual, dialogue, and interactive modality. To offset over-reliance on recall, we consider providing participants with a way to review the explanation in future studies. Lastly, we noticed a bias towards higher education in our participant sample, with 22 participants having a university degree. However, we aimed to collect different perspectives by speaking to employees from the local employment agency, job-seekers and individuals with different educational backgrounds.

\section{Conclusion}

In this paper, we inductively and deductively analyse participants' understanding of an algorithmic decision-making system after they received one of three explanation modalities (textual, dialogue, and interactive). We find that all of the six "facets of understanding"~\cite{wiggins_understanding_2005} (explain, interpret, apply, empathise, take perspective, self-reflect) emerge in participant responses throughout the study, with some participants expressing high understanding by combining multiple facets at once. We argue that incorporating theories from the learning sciences can significantly improve the design of ADM explanations by adapting them to the underlying thought processes of learning and understanding in individuals. 
We further highlight the "dialogue" explanation modality as a valid alternative to convey information and gather in-depth insights on how participants understand and contextualise explanations. Lastly, while we observe that most participants are able to articulate a fairness assessment of the explained ADM system and that a more pronounced understanding supports this articulation, it also becomes evident that participants have general difficulties considering the meaning of "fairness" in the context of algorithmic systems. We posit that letting stakeholders independently assess algorithmic systems in terms of ethical values such as fairness, accountability, or transparency, could require an additional explanation of how to apply these values in an algorithmic context, in addition to increasing stakeholders' understanding of the algorithm itself.

\begin{acks}
This work has been funded by the Vienna Science and Technology Fund (WWTF) [10.47379/ICT20058] as well as [10.47379/ICT20065].
\end{acks}


\bibliographystyle{ACM-Reference-Format}
\bibliography{references}




\end{document}